\documentclass[11pt]{article}
\usepackage{graphicx}
\usepackage{amsmath}
\usepackage{epsf}
\usepackage{epsfig}
\usepackage{amssymb}
\usepackage{epstopdf}
\usepackage{latexsym}
\usepackage{subfigure}
\begin{document}
\begin{titlepage}
$\mbox{ }$
\vspace{.1cm}
\begin{center}
\vspace{.5cm}
{\bf\Large Electroweak Constraints on Warped Geometry in Five Dimensions and Beyond}\\[.3cm]
\vspace{1cm}
Paul R.~Archer$^{a,}$\footnote{p.archer@sussex.ac.uk}
and 
Stephan J.~Huber$^{a,}$\footnote{s.huber@sussex.ac.uk} \\ 
\vspace{1cm} {\em  
$^a$Department of Physics \& Astronomy, University of Sussex, Brighton
BN1 9QH, UK }\\[.2cm]

\end{center}
\bigskip\noindent
\vspace{1.cm}

\begin{abstract}
Here we consider the tree level corrections to electroweak (EW) observables from standard model (SM) particles propagating in generic warped extra dimensions. The scale of these corrections is found to be dominated by three parameters, the Kaluza-Klein (KK) mass scale, the relative coupling of the KK gauge fields to the Higgs and the relative coupling of the KK gauge fields to fermion zero modes. It is found that 5D spaces that resolve the hierarchy problem through warping typically have large gauge-Higgs coupling. It is also found in $D>5$ where the additional dimensions are warped the relative gauge-Higgs coupling scales as a function of the warp factor. If the warp factor of the additional spaces is contracting towards the IR brane, both the relative gauge-Higgs coupling and resulting EW corrections will be large. Conversely EW constraints could be reduced by finding a space where the additional dimension's warp factor is increasing towards the IR brane. We demonstrate that the Klebanov Strassler solution belongs to the former of these possibilities.         
\end{abstract}
\end{titlepage}

\section{Introduction}
For over a decade now, much interest has been paid to the possibility of warped extra dimensions, principally because they offer a non-supersymmetric resolution to the hierarchy problem \cite{Randall:1999ee}. Since its original proposal the Randall and Sundrum (RS) model has been extended to allow standard model (SM) particles to propagate into the bulk \cite{Davoudiasl:1999tf}\cite{Pomarol:1999ad}\cite{Grossman:1999ra}\cite{Chang:1999nh}, offering explanations for the flavour hierarchy \cite{Huber:2000ie}\cite{Gherghetta:2000qt} as well as the suppression of flavour changing neutral currents \cite{Gherghetta:2000qt}\cite{Agashe:2004cp}. The RS model has now grown into a large and varied framework for building models of new physics. However the vast majority of this work is restricted to considering the original 5D slice of AdS space. While the RS model may be considered as a toy model, it does have motivations from AdS/CFT, where it has been proposed that it may arise as the dual of $\mathcal{N}=4$ Super Yang Mills with a IR/UV cut off \cite{ArkaniHamed:2000ds}. Again since the original proposal of AdS/CFT \cite{Maldacena:1997re}, many other duals have been found, for example \cite{Klebanov:2000nc}\cite{Klebanov:2000hb}. This and the partial success in using AdS/CFT to model QCD suggests there may be many more out there and that it is perhaps unwise when model building is restricted to straight AdS$_5\times S_5$ or for that matter just AdS$_5$.

Further motivation for looking beyond pure AdS$_5$ comes from studying electroweak (EW) constraints: With SM fields in the bulk, the Kaluza-Klein (KK) scale is pushed to at least 10 TeV \cite{Davoudiasl:1999tf}\cite{Csaki:2002gy}. Well known mechanisms for relaxing these bounds involve large brane kinetic terms  \cite{carena-2003-68} or a left-right symmetric extension of the SM gauge group \cite{Agashe:2003zs}. In both cases the KK scale can be brought down to a few TeV. It is natural to ask if this picture remains unchanged if the RS framework is generalized. Obviously, this issue
is of direct relevance to the LHC experiment.

With the LHC now in operation it is of the upmost importance to keep models as generic as possible so as to not misinterpret any signals. For example a $W^{\prime}$ or $Z^{\prime}$ boson is one of the feasible early discoveries and clearly it would be useful to know if we can ascribe them to extra dimensions. With this in mind here we consider the electroweak (EW) constraints arising from a generic D-dimensional warped background. Work has already been done looking at AdS$_5\times\mathcal{M}^\delta$ \cite{Davoudiasl:2002wz}\cite{Davoudiasl:2008qm} where little deviation from the 5D RS model was found. It has also been shown for generic 5D models, without custodial symmetry, that resolving the hierarchy problem one does typically get quite large EW constraints \cite{Delgado:2007ne}. Also work has been done going beyond $AdS_5$ with deformed conifolds \cite{McGuirk:2007er} and on $AdS_D$ \cite{McDonald:2009hf}\cite{McDonald:2009md}. 

In this work we generalize the work described above and study in detail electroweak constraints in modifications of the original RS geometry, both in five and higher dimensions. Here we consider a D-dimensional space warped with respect to a single extra dimension and bounded by two branes (IR and UV). Electroweak symmetry is broken by a Higgs localised on the IR brane while the SM fermions and gauge fields are free to propagate in the bulk. In section 2 we describe this model in more detail an consider the bulk gauge and fermion fields.

Allowing the SM particles to propagate in the bulk leads to a tree level contributions to the EW observables. In section 3 these corrections are computed and it is found that they are largely dependent on three parameters, the coupling between the KK gauge fields and the Higgs, the fermion zero mode coupling to the KK gauge fields and the KK mass scale. Taking a best case scenario in which the fermions are localised away from the KK gauge fields and hence the relative gauge-fermion coupling is suppressed, the lower bound on the KK scale is then determined by the size of the relative gauge-Higgs coupling.    

In section 4 we move on to look at the relative gauge-Higgs coupling. Firstly in 5D where it is found a space that resolves the hierarchy problem through warping and has a small coupling would be quite contrived. Hence we agree with \cite{Delgado:2007ne} in saying that large EW constraints appear to be a generic feature of 5D warped spaces. On the other hand in D dimensions it is found that where the additional dimensions are also being warped the volume enhancement/suppression of the couplings are different for the KK modes than for the zero modes. The result is the relative gauge-Higgs couplings scale as a function of the overall warp factor. We finish by testing this result by computing the relative coupling for gauge fields propagating in a 10D solution of type IIb supergravity, the Klebanov-Strassler solution \cite{Klebanov:2000hb}, and find them to be very large. In section 5 we conclude.   

It should be stressed that our interests are in a low energy effective theory and hence, for example, we do not consider any back-reaction effects or other high energy effects. It also should be pointed out that although we mostly focus on a potentially large gauge-Higgs coupling, all matter on the IR brane will couple in a similarly strong way.

\section{General Framework}
\subsection{The Metric}
Here we consider a D-dimensional warped space of the form
\begin{equation}
\label{Metric}
ds^2=a^2(r)\eta_{\mu\nu}dx^\mu dx^\nu-b^2(r)dr^2-c^2(r)d\Omega_\delta^2,
\end{equation}
where $\eta_{\mu\nu}=\mbox{diag}(+---)$ is the 4D Minkowski metric, and $r$ parametrizes the warped 5th dimension. We also allow for an additional internal manifold that is described by $d\Omega_\delta^2=\gamma_{ij}d\phi^id\phi^j$, with $i,j$ running from $1\dots \delta$. The total spacetime dimension is $D=4+1+\delta$. Note that it is always possible to set $b=1$ with a coordinate transformation $r\rightarrow\tilde{r}=\int_c^r b(\hat{r})d\hat{r}$, however including $b$ allows us to analytically express a greater range of spaces. 

As with other warped scenarios we cut the space off with two branes. The UV brane is defined to be located such that $a(r_{\rm{uv}})=1$. On the IR brane at  $r_{\rm{ir}}$ the Higgs is localised. The 4D effective Higgs mass will then be suppressed down from its fundamental value 
$m_{\rm{fund}}$ by
\begin{equation}
\label{HiggsMass}
m_{\rm{4D}}^2=a^2(r_{\rm{ir}})m_{\rm{fund}}^2
\end{equation} 
while the 4D Planck mass will scale, relative to its fundamental value $M_{\rm{fund}}$, as 
\begin{equation}
\label{PlankMass}
M_{\rm{P}}^2\sim\int d^{\delta+1}x\; a^2bc^\delta \sqrt{\gamma}\;M_{\rm{Fund}}^{\delta +3}.
\end{equation}
The now well known extra dimensional resolution to the gauge hierarchy problem is to suppose that the fundamental Higgs mass is of the same order as the fundamental Planck mass. In the ADD model volume effects are used to suppress the fundamental Planck scale down to the EW scale \cite{ArkaniHamed:1998rs}\cite{Antoniadis:1998ig}. On the other hand in the RS model these volume effects are only of order one, while the Higgs mass is exponentially suppressed by gravitational red shifting.  Both effects can be at work at the same time, so when considering warped extra dimensions the amount of warping required to resolve the hierarchy problem is dependent on the volume of the space. Here we parameterise this in terms of the warp factor defined to be $\Omega\equiv a(r_{\rm{ir}})^{-1}$ (not to be confused with the metric of the internal manifold) and note that without the volume effect, a warp factor of $\Omega\sim 10^{15}$ is required to solve the hierarchy problem. Alternatively, one could postpone the full resolution of the hierarchy problem to a higher energy and reduce the UV cut off to a scale much lower than the Planck scale, as in the little RS model. This would essentially amount to a reduction in the required value of $\Omega$ and as in the 5D case \cite{Davoudiasl:2008hx}, one would anticipate this reducing the EW constraints.

\subsection{Bulk Gauge Fields}
We now compute the low energy effective action of a gauge field propagating in the bulk (\ref{Metric}) described by
\begin{equation}
\label{ }
S=\int d^Dx\; \sqrt{-G}\left [-\frac{1}{4}A_{MN}A^{MN}\right ],
\end{equation} 
where $A_{MN}$ is the higher-dimensional field strength tensor. Expanding and integrating by parts gives
\begin{displaymath}
S=\int d^Dx\; \bigg [ -\frac{1}{4}bc^\delta \sqrt{\gamma} A_{\mu\nu}A^{\mu\nu}-\frac{1}{2}\partial_r\left (a^2b^{-1}c^\delta \sqrt{\gamma} \partial_r A_\mu\right )A^\mu\end{displaymath}
\begin{displaymath}
\hspace{3cm}-\frac{1}{2}\sum_{\phi_i,\phi_j}\partial_{\phi_i}\left (a^2bc^{\delta-2}\sqrt{\gamma}\gamma^{ij}\partial_{\phi_j}A_\mu\right )A^\mu+\dots \bigg ]
\end{displaymath}
where we have assumed that $A_\mu$ satisfies either Neumann or Dirichlet boundary conditions (BCs). In practice we want to describe SM particles with the zero-mode which requires Neumann BCs. The dots represent the remaining terms which are either higher order in $A_\mu$ or contain the  higher-dimensional  components of the gauge field. While it is always possible to gauge away one of these components, the four dimensional theory would still be left with $\delta$ effective scalar fields. Since we do not observe any low mass charged scalar fields it is necessary to fix the boundary conditions such that these field do not gain zero modes. In the following we will assume that this possible. Further still imposing Dirichlet boundary conditions on the gauge-scalars will ensure they largely decouple from the electroweak sector.

We decompose the field as
\begin{equation}
\label{ }
A_\mu=\sum_{n}A_\mu^{(n)}(x^\mu)f_n(r)\Theta_n(\phi_1,\dots,\phi_\delta)
\end{equation}
such that
\begin{equation}
\label{gaugeortho}
\int d^{1+\delta}x\; bc^\delta\sqrt{\gamma}f_nf_m\Theta_n\Theta_m=\delta_{nm},
\end{equation}
then the low energy effective action is given by
\begin{equation}
S_{eff.}=\int d^4x\;\sum_n\left [-\frac{1}{4}A_{\mu\nu}^{(n)}A^{\mu\nu}_{(n)}+\frac{1}{2}m_n^2A_\mu^{(n)}A^{\mu}_{(n)}\dots\right ]
\end{equation} 
where $m_n$ is given by
\begin{equation}
\label{gaugeEQN}
f_n^{\prime\prime}+\frac{(a^2b^{-1}c^\delta)^{\prime}}{(a^2b^{-1}c^\delta)}f_n^{\prime}-\frac{b^2}{c^2}\alpha_nf_n+\frac{b^2}{a^2}m_n^2f_n=0.
\end{equation}
Here $^{\prime}$ denotes the derivative with respect to $r$, and $\alpha_n$ is the eigenvalue of the Laplacian operator 
\begin{displaymath}
-\frac{1}{\sqrt{\gamma}}\partial_{\phi_i}(\sqrt{\gamma}\gamma^{ij}\partial_{\phi_j}\Theta_n)=\alpha_n\Theta_n.
\end{displaymath}
Note that there is now a degeneracy in the KK spectrum corresponding to the ``harmonic" modes, i.e.~modes that are related  to excitations in the $\delta$ additional dimensions. However,  generally $\alpha_n\geqslant 0$ and so the mass of the first ``harmonic" mode will always be greater than the corresponding ``non-harmonic" mode. 

\subsection{Fermions}

In 5D it is well known that by choosing a suitable 5D Dirac mass the bulk fermions zero modes can be localised towards the UV brane \cite{Grossman:1999ra}, resulting in suppressed couplings with the KK gauge fields. The low energy, 4D effective chiral theory is achieved by an appropriate choice of boundary conditions. However these results do not straight forwardly apply to D$>5$ dimensions. In even dimensions, where the fermions can be described by a $2^{\frac{D}{2}}$-component Dirac representation, the introduction of a Dirac mass term would only be possible if the higher dimensional theory was non chiral. This could potentially lead to more than one zero mode of the $4$D $2$-component Weyl spinor. In odd dimensions where the Dirac representation would be a $2^{\frac{D+1}{2}}$ vectorial quantity, Dirac masses are allowed. In both cases it is probably possible to choose appropriate orbifolding to ensure the correct 4D chiral theory \cite{McDonald:2009hf}\cite{Burdman:2005sr}. However it is not straight forward to do so while still working with a generic dimensionality.

Here we consider a `best case scenario' in which the fermions are free to propagate in the bulk and could potentially be localised away from the gauge fields. We assume that included in the higher dimensional fermion is a 2-component Weyl spinor($\psi_{\rm{L}}$) and that due to appropriate boundary conditions, only this part of the fermion gains a zero mode. The alternative to this scenario is of course to localise the fermions on the IR brane in which case the relative gauge-fermion coupling would be the same as the relative gauge-Higgs coupling (see section 3). 

As we shall see the tree level EW constraints are only dependent on the zero mode of $\psi_{\rm{L}}$ and even then they are in practice dominated by the gauge-Higgs coupling. So here we leave it to future work to investigate the full phenomenological implications of bulk fermions and instead rather schematically refer to the KK decomposition of $\psi_{\rm{L}}$,
\begin{equation}
\psi_{\rm{L}}=\sum_n\psi_{\rm{L}}^{(n)}(x^\mu)f_{\rm{L}}^{(n)}(r)\Theta_{\rm{L}}^{(n)}(\phi_1,\dots,\phi_\delta)
\end{equation}
with the orthogonality relation
\begin{equation}
\int d^{\delta+1}x a^3bc^\delta\sqrt{\gamma} f_{\rm{L}}^{(n)}f_{\rm{L}}^{(m)}\Theta_{\rm{L}}^{(n)}\Theta_{\rm{L}}^{(m)}=\delta_{nm}
\end{equation}
chosen such that the kinetic term of the Lagrangian  is orthogonal.

\section{The Electroweak Sector}
We assume that the electroweak symmetry is broken by a Higgs boson and that in order to resolve the gauge hierarchy problem the Higgs must be localised close to the IR tip of the space. This then presents two possibilities, firstly that the Higgs is localised to a 3-brane and hence not free to propagate in the internal manifold. So the EW sector would be described by
\begin{eqnarray}
S=\int d^Dx \sqrt{-G}\Big [-\frac{1}{4}A_{MN}^aA^{MN\;a}-\frac{1}{4}B_{MN}B^{MN}+\sum_{\Psi}\left (i\bar{\Psi}E^M_A\Gamma^A\nabla_M\Psi-M\bar{\Psi}\Psi\right )\Big ]\nonumber\\
+\int d^4x\sqrt{-g_{\rm{ir}}}\left [|D_\mu\Phi|^2+V(\Phi)\right ]\label{SMaction1},
\label{5Daction}
\end{eqnarray}

where $g_{\rm ir}^{\mu\nu}$ is the induced metric $G^{\mu\nu}(x^\mu,r_{\rm{ir}},\phi_{\rm{ir}})$. $A_{MN}$ and $B_{MN}$ are the field strength tensors for the $SU(2)$ and $U(1)$ gauge fields. While $E^M_A$ is the vielbein defined such that $E_M^A\eta_{AB}E_N^B=G_{MN}$. As mentioned earlier we include a bulk Dirac mass ($M$) assuming a best case scenario in which the fermions are localised away from the gauge fields.
 
The second option is that the Higgs is localised only in the one warped co-ordinate, $r$, and is free to propagate in the internal manifold. Hence it would be described by
\begin{eqnarray}
S=\int d^Dx \sqrt{-G}\Big\{-\frac{1}{4}A_{MN}^aA^{MN\;a}-\frac{1}{4}B_{MN}B^{MN}+\sum_{\Psi}\left (i\bar{\Psi}E_A^M\Gamma^A\nabla_M\Psi-M\bar{\Psi}\Psi\right )\nonumber\\
+\frac{\delta(r-r_{\rm{ir}})}{b}\Big [|D_\mu\Phi|^2 +V(\Phi)\Big ]\Big\}\label{SMaction2}.
\end{eqnarray}
In computing the low energy effective action we are faced with two equivalent options, either to carry out the KK decomposition pre spontaneous symmetry breaking (SSB) or post SSB. Post SSB the boundary mass term will induce a non trivial BC but the resulting terms of KK modes will all be orthogonal. On the other hand pre SSB one can use the original Neumann or Dirichlet BCs but the Higgs terms will mix the KK modes, and hence the mass matrix will gain off diagonal terms. It is the latter option that we use here. Let us return to our two possible Higgs localisations. The first terms of  the actions can be expanded as described in the previous section, while the covariant derivative is now given as
\begin{equation}
\label{ }
D_\mu=\partial_\mu+\sum_n\left (-igf_n\Theta_nA_\mu^{a(n)}\tau^a-iYg^{\prime}f_n\Theta_nB_\mu^{(n)}\right ).
\end{equation}
The EW symmetry is broken by $\Phi\rightarrow\frac{a(r_{\rm{ir}}^{-1})}{\sqrt{2}}\left (\begin{array}{ c}0\\v+H\end{array}\right )$. We perform the usual field redefinitions
\begin{eqnarray}
A_\mu^{1(n)}=\frac{1}{\sqrt{2}}\left (W_\mu^{+(n)}+W_\mu^{-(n)}\right )\qquad A_\mu^{2(n)}=\frac{i}{\sqrt{2}}\left (W_\mu^{+(n)}-W_\mu^{-(n)}\right )\nonumber\\
A_\mu^{3(n)}=cZ_\mu^{(n)}+sA_\mu^{(n)}\qquad B_\mu^{(n)}=cA_\mu^{(n)}-sZ_\mu^{(n)}\nonumber
\end{eqnarray}

where
\begin{equation}
c\equiv\frac{g}{\sqrt{g^2+g^{\prime 2}}}\qquad s\equiv\frac{g^{\prime}}{\sqrt{g^2+g^{\prime 2}}} .
\end{equation} 
The gauge field mass matrices can then be computed from the Higgs kinetic term. We can now see the difference between the two models. When the Higgs is free to propagate in the internal manifold (\ref{SMaction2}) the `harmonic modes', orthogonal under (\ref{gaugeortho}), will decouple from the gauge-zero mode mass term and hence
  \begin{equation}
\label{ }
|D_\mu\Phi|^2\supset\sum_{n,m}\frac{g^2v^2}{4}f_nf_mW_\mu^{+(n)}W^{-(m)\mu}+\frac{(g^2+g^{\prime 2})v^2}{8}f_nf_mZ_\mu^{(n)}Z^{\mu (m)}
\end{equation}
where $f_n=f_n(r_{\rm{ir}})$. On the other hand, where the Higgs is localised to a 3-brane (\ref{SMaction1}) the orthogonality relations cannot generally be applied and hence  
 \begin{equation}
\label{ }
|D_\mu\Phi|^2\supset\sum_{n,m}\frac{g^2v^2}{4}f_n\Theta_nf_m\Theta_mW_\mu^{+(n)}W^{-(m)\mu}+\frac{(g^2+g^{\prime 2})v^2}{8}f_n\Theta_nf_m\Theta_mZ_\mu^{(n)}Z^{\mu (m)}
\end{equation}
Bearing in mind that for a given $m_n$, $\Theta_n$ will typically contain a sum over allowed degeneracies and hence $\Theta_n\Theta_m$ will typically be a sub matrix. Hence when the Higgs is localised on a 3-brane 
there are more modes mixing with the zero mode
in the mass matrix. However, the EW corrections will be largely dominated by the first KK mode, which, as already mentioned, will always correspond to a $\alpha_n=0$ non-harmonic mode.

\subsection{Electroweak Observables}
It is well known that extra dimensions contribute corrections to EWOs at tree level. Such corrections are suppressed by the KK scale and so a lower bound on the KK scale can be computed. Often these corrections are parameterised in terms of S and T parameters  \cite{Peskin:1991sw}. However, here we think it is more transparent to compute the corrections to the EWO's explicitly.  Before one can compute the corrections to a given observable it is necessary to fix the input parameters, $g$, $g^{\prime}$ and $v$. Ideally this would be done through a $\chi^2$ test. However, for the purpose of this study it is sufficient to fix these parameters by comparison with the three most precisely observed quantities \cite{Amsler:2008zzb},
\begin{equation}
\label{ }
\hat{\alpha}(M_Z)^{-1}=127.925\pm0.016\quad G_f=1.166367(5)\times10^{-5} \;\mbox{GeV}^{-2}\quad \hat{M}_Z=91.1876\pm0.0021\; \mbox{GeV}.
\end{equation}
 Before going any further it is useful to note that the W and Z boson KK masses are given as eigenvalues of the matrices
 \begin{equation}
\label{massmatrix}
(M_W^2)_{mn}=m_n^2\delta_{mn}+\frac{g^2v^2}{4}f_nf_m\qquad(M_Z^2)_{mn}=m_n^2\delta_{mn}+\frac{(g^2+g^{\prime 2})v^2}{4}f_nf_m.
\end{equation} 
When the Higgs is localised to a 3-brane, the same expression arises, but with $\Theta_n$'s included. We also define the gauge fermion coupling
 \begin{displaymath}
f_\psi^{(l,n,m)}\equiv\int\, d^{\delta+1}x\, ba^3c^\delta\sqrt{\gamma}\,f_{L}^{(l)}\Theta_L^{(l)}\,f_{n}\Theta_n\,f_{L}^{(m)}\Theta_L^{(m)}.
\end{displaymath} 
It is worth pointing out that the bulk fermion zero mode will be constant in the $\phi$ direction and hence will decouple from the gauge fields harmonic modes using (\ref{gaugeortho}). Hence the gauge harmonic modes would not contribute to most SM tree level processes involving fermions. On the otherhand if the fermions were localised to the brane the gauge harmonic modes would be relevant. 

\begin{figure}
\begin{center}
\includegraphics[width=6in] {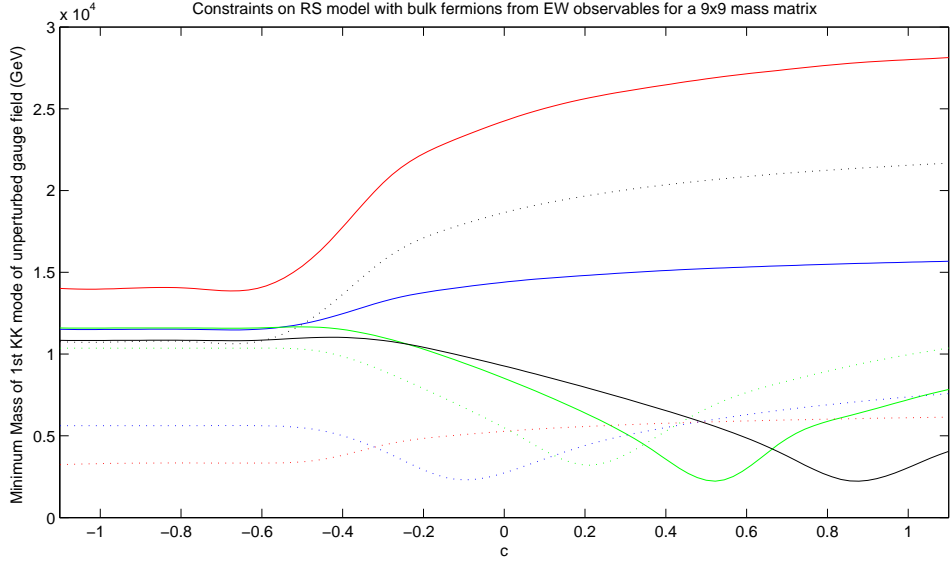}
\caption{\footnotesize The constraints on the 5D RS model with the SM propagating in the bulk. Plotted is the lower bound on the first KK gauge mass ($\sim2.45M_{KK}$) arising from comparison with experimental error on the EWO's; $S_Z^2$ (Red line), $M_W$ (blue line), $\Gamma_Z$ (green line), $\Gamma_{\mbox{had}}$ (black line),  $R_e$ (blue dots), $\Gamma_{\mbox{inv}}$ (red dots), $\Gamma_{l^+l^-}$ (green dots) and $A_e$ (black dots).  On the horizontal axis is plotted the 5D Dirac mass, $c=\frac{M}{R}$ with $c<-0.5$($>-0.5$) meaning the fermions are localized towards the UV (IR) brane.}
\label{RScouplingfig}
\end{center}
\end{figure}

Likewise if we were to consider models where the fermions are localised to 3-branes then $f_\psi^{(n)}=f_n(r_{\rm{ir}})\Theta_n(\phi_{\rm{ir}})$, i.e.~proportional  to the gauge Higgs coupling. Also it should be noted that in order get the actual gauge fermion couplings, one would need to rotate to a basis where the gauge boson mass matrix is diagonal, i.e.~if $U_Z$ is a unitary matrix such that $U_Z(M_Z^2)U_Z^{-1}$ is diagonal then the   Z boson-fermion coupling would be given as $f_Z^{(l,m,k)}=(U_Z^{-1})^{mn}f_\psi^{(l,n,k)}$. With these definitions our input observables are then given at tree level as
 \begin{eqnarray}
\hat{M}_Z^2=\left (U_Z(M_Z^2)U_Z^{-1}\right )_{00}\label{MZobs}\\
\sqrt{4\pi\alpha(M_Z)}=\frac{gg^{\prime}}{\sqrt{g^2+g^{\prime2}}}f_\psi^{(0,0,0)}\label{alpha}\\
4\sqrt{2}G_f=g^2(f_\psi^{(0,m,0)})^\dag\,(M_W^2)_{mn}^{-1}f_\psi^{(0,n,0)}\label{Gf}
\end{eqnarray}
The equations can then be solved for $g$, $g^{\prime}$ and $v$ and the remaining EWO's can  be computed and compared with experimental results. The analysis used here aims at determining the minimal KK scale such that the deviation between the tree level EWO and the tree level standard model predicted value was within $2\sigma$ of the experimental result. Since we are just working to tree level we compare to LEP 1 Z pole data. What is typically found is that the tightest constraint comes from the weak mixing angle $S_Z^2$. As way of an example the results for the 5D Randall and Sundrum model are shown in figure \ref{RScouplingfig}. These results give comparable constraints to existing studies \cite{Huber:2001gw}\cite{Csaki:2002gy}\cite{carena-2003-68}\cite{Davoudiasl:1999tf}. However before considering specific models, in the interests of generality, it is useful to look at how these EW corrections depend on generic couplings. 

\subsection{Approximate Expressions for the EW Corrections}
To proceed we have to diagonalise the mass matrices. Here we use a perturbative approach outlined in \cite{Goertz:2008vr}.  In any realistic model the off diagonal terms will be relatively small. So given a matrix of the form
 \begin{displaymath}
M=\left(\begin{array}{ccccc}A_{1} & B_{12} & B_{13} & \cdots & B_{1N} \\B_{12} & A_2 & B_{23} & \cdots &  \\B_{13} & B_{23} & A_3 &  &  \\\vdots &\vdots  &  & \ddots &  \\B_{1N} &  &  &  & A_N\end{array}\right)
\end{displaymath}
When $A\gg B$ the eigenvalues are approximately $\lambda_n\,\approx A_n-\sum_{i\neq n}\frac{B_{ni}^2}{A_i-A_n}$ and $M$ can be diagonalised by
\begin{equation}
\label{Uinv}
U^{-1}\approx\left(\begin{array}{cccc}1 & -\frac{B_{12}}{A_2-A_1} & -\frac{B_{13}}{A_3-A_1} & \cdots \\\frac{B_{12}}{A_2-A_1} & 1 & -\frac{B_{23}}{A_3-A_2} &  \\\frac{B_{13}}{A_3-A_1} & \frac{B_{23}}{A_3-A_2} & 1 &  \\\vdots & \vdots &  & \ddots\end{array}\right)
\end{equation} 
Before we apply this to our case it is useful to define the relative gauge-Higgs coupling and the relative gauge fermion coupling to be
 \begin{displaymath}
F_n\equiv\frac{f_{n}(r_{\rm{ir}})}{f_{0}(r_{\rm{ir}})}\quad \mbox{or}\quad F_n\equiv\frac{f_{n}(r_{\rm{ir}})\Theta_n(\phi_{\rm{ir}})}{f_{0}(r_{\rm{ir}})\Theta_0(\phi_{\rm{ir}})}\quad \mbox{and}\quad F_\psi^{(n)}\equiv\frac{f_\psi^{(0,n,0)}}{f_\psi^{(0,0,0)}}.
\end{displaymath}
Clearly $F_n$ may include a sum over the degenerate harmonic modes. We also define the SM W and Z masses
\begin{displaymath}
m_w^2\equiv\frac{g^2v^2}{4}f_{0}(r_{\rm{ir}})^2\quad\mbox{and}\quad m_z^2\equiv\frac{(g+g^{\prime\,2})v^2}{4}{f_{0}(r_{\rm{ir}})}^2.
\end{displaymath}
We start by computing the correction to the weak mixing angle which at tree level can be related to our input observables by
\begin{displaymath}
s_Z^2=\frac{\pi\alpha}{\sqrt{2}G_f}\frac{f_\psi^{(0,m,0)}(M_W^2)^{-1}_{nm}f_\psi^{(0,n,0)}}{(f_\psi^{(0,0,0)})^2}.
\end{displaymath}
The inverse of the W mass matrix has the relatively simple form
\begin{displaymath}
(M_W^2)^{-1}=\left(\begin{array}{ccccc}(M_W^2)^{-1}_{00} & -\frac{F_1}{m_1^2} & -\frac{F_2}{m_2^2} & \cdots & -\frac{F_n}{m_n^2} \\-\frac{F_1}{m_1^2} & \frac{1}{m_1^2} & 0 & \cdots &  \\-\frac{F_2}{m_2^2} & 0 & \frac{1}{m_2^2} &  &  \\\vdots & \vdots &  & \ddots &  \\-\frac{F_n}{m_n^2}&  &  &  & \frac{1}{m_n^2}\end{array}\right)
\end{displaymath}
where
\begin{displaymath}
(M_W^2)^{-1}_{00}=\frac{1}{m_w^2}+\sum_{n=1}\frac{F_n^2}{m_n^2}
\end{displaymath}
and hence
\begin{equation}
\label{szapprox1}
s_Z^2=\frac{\pi\alpha}{\sqrt{2}G_f}\left (\frac{1}{m_w^2}+\sum_{n=1}\frac{\left(F_n-F_\psi^{(n)}\right )^2}{m_n^2}\right ).
\end{equation}
We can solve for $m_w^2$ by noting that $m_w^2=m_z^2(1-s_Z^2)$ and using the above relation that 
\begin{displaymath}
\hat{M}^2_Z\approx m_z^2\left (1-\sum_{n=1}\frac{m_z^2F_n^2}{m_n^2}+\mathcal{O}(m_n^{-4})\right )
\end{displaymath}
If we now define the predicted weak mixing angles to be
\begin{equation}
\label{spcp}
s_p^2=\frac{1}{2}\left (1-\sqrt{1-\frac{4\pi\alpha}{\sqrt{2}G_f\hat{M}_Z^2}}\right ),\qquad c_p^2=\frac{1}{2}\left (1+\sqrt{1-\frac{4\pi\alpha}{\sqrt{2}G_f\hat{M}_Z^2}}\right ).
\end{equation}
Then after a little algebra one arrives at the result
\begin{equation}
\label{ }
s_Z^2\approx s_p^2\left (1-\frac{c_p^2}{c_p^2-s_p^2}\sum_{n=1}\left [\frac{m_z^2F_n^2}{m_n^2}-\frac{m_w^2\left(F_n-F_\psi^{(n)}\right )^2}{m_n^2}\right ]+\mathcal{O}(m_n^{-4})\right ).
\end{equation}
Likewise the correction to the W mass can be computed using an analogous method
\begin{equation}
\label{ }
M_W^2\approx c_p^2\hat{M}_Z^2\left (1+\sum_{n=1}\left [\frac{(m_z^2-m_w^2)F_n^2}{m_n^2}\right ]+\frac{s_p^2}{c_p^2-s_p^2}\sum_{n=1}\left [\frac{m_z^2F_n^2}{m_n^2}-\frac{m_w^2\left(F_n-F_\psi^{(n)}\right )^2}{m_n^2}\right ]+\mathcal{O}(m_n^{-4})\right ).
\end{equation}
Note that provided $F_\psi^{(n)}<F_n$, the W mass is always shifted in the correct direction with regard to recent precision measurements at the Tevatron \cite{Aaltonen:2007ps}.  Another quantity of importance for EW observables is the extent to which the Z  coupling is perturbed by the boundary mass,  which using (\ref{Uinv}) is given by
 \begin{equation}
\label{ }
\sqrt{g^2+g^{\prime\;2}}f_Z^{(0)}\approx \sqrt{g^2+g^{\prime\;2}} f_\psi^{(0)}\left (1-\sum_{n=1}\frac{m_z^2F_nF_\psi^{(n)}}{m_n^2}+\mathcal{O}(m_n^{-4})\right ).
\end{equation}    
This shows that the size of the EW corrections is determined by three quantities, the relative gauge-Higgs coupling $F_n$, the relative gauge-fermion coupling $F_\psi^{(n)}$ and the of course the KK mass scale. In our best case scenario where the $F_\psi^{(n)}$ can be suppressed by, for example localising towards the UV brane, then the lower bound on the KK scale arising from EW corrections is largely determined by the scale of $F_n$. For example the 5D RS model with $F_n\approx8.3$ (for $\Omega=10^{15}$) typically has much larger constraints than that of universal flat extra dimensions with $F_n\approx \sqrt{2}$.  

A well known cure for these large constraints is to impose a custodial $SU(2)_R\times SU(2)_L$ symmetry on the bulk which essentially fixes $\rho\equiv\frac{m_w^2}{c^2m_z^2}$ to be 1 \cite{Agashe:2003zs}. The result of this is that when the additional $SU(2)_R$ gauge field is included in the $\frac{(m_z^2-m_w^2)F_n^2}{m_n^2}$ term, it cancels to zero and hence the EW constraints would be linearly dependent on $F_n$ rather than quadratically. But none the less a large value of $F_n$  will mean the KK mass scale must also be large. The remainder of this paper will be focused on determining which geometries result in large values of $F_n$.

\section{The Relative Gauge-Higgs Coupling}
The relative gauge-Higgs coupling is given by
\begin{equation}
\label{Fn}
F_n=\frac{\sqrt{\int bc^\delta\sqrt{\gamma}\,d^{\delta+1}x}\, f_n(r_{\rm{ir}})\Theta_n(\phi_{\rm{ir}})}{\sqrt{\int bc^\delta\sqrt{\gamma}f_n^2\Theta_n^2\,d^{\delta+1}x}}.
\end{equation}  
However even in the case when the Higgs is localised to a 3-brane the EW constraints will be dominated by the first KK mode which will be a non-harmonic mode where $\Theta_n$ is a constant. So here we are mostly interested in
 \begin{equation}
 \label{fnsmall}
F_n=\frac{\sqrt{\int bc^\delta\,dr}\, f_n(r_{\rm{ir}})}{\sqrt{\int bc^\delta f_n^2\,dr}}.
\end{equation}
Before considering $D>5$ it is worth briefly looking at 5D.

\subsection{Are Large $F_n$ Values a Generic Feature of Backgrounds that Resolve the Hierarchy Problem?}
Here we look at the question (first considered in \cite{Delgado:2007ne}), can a background be found that resolves the hierarchy problem and still has small EW constraints. If we start by rewriting the 5D version of (\ref{gaugeEQN}), i.e. with $\delta=\alpha_n=0$ as
\begin{displaymath}
\left (a^2b^{-1}\exp\left [\int_c^r\frac{f^{\prime\prime}(\tilde{r})}{f^{\prime}(\tilde{r})}d\tilde{r}\right ]\right )^{\prime}=-bm_n^2\frac{f}{f^{\prime}}\exp\left [\int_c^r\frac{f^{\prime\prime}(\tilde{r})}{f^{\prime}(\tilde{r})}d\tilde{r}\right ].
\end{displaymath}
This can be solved for $a(r)$
\begin{equation}
\label{a2}
a^2(r)=\frac{-Km_n^2b(r)\int_c^rb(\tilde{r})f_n(\tilde{r})d\tilde{r}}{f_n^{\prime}(r)},
\end{equation}
where $K$ is a constant of integration. In order to determine whether a space resolves the hierarchy problem or not, we are interested in the warp factor defined to be $\Omega=a^{-1}(r_{\rm{ir}})$. But clearly under Neumann BC's (\ref{a2}) is ill defined. So evaluating (\ref{a2}) at the boundaries
\begin{displaymath}
a^2(r_{\rm{ir/uv}})=\lim_{\delta\rightarrow 0}a^2(r_{\rm{ir/uv}}+\delta)=\frac{-Km_n^2b^2(r_{\rm{ir/uv}})f_n(r_{\rm{ir/uv}})}{f_n^{\prime\prime}(r_{\rm{ir/uv}})}
\end{displaymath}
and fixing the the UV brane at $a(r_{\rm{uv}})=1$ then we arrive at the expression
\begin{equation}
\label{HierCon}
\Omega^2=\frac{b^2(r_{\rm{uv}})f_n(r_{\rm{uv}})f_n^{\prime\prime}(r_{\rm{ir}})}{b^2(r_{\rm{ir}})f_n(r_{\rm{ir}})f_n^{\prime\prime}(r_{\rm{uv}})}.
\end{equation}
That is to say the wave function of a gauge field propagating in a space with a warp factor $\Omega$ would satisfy this condition. Typically one would require $\Omega$ to be large of order $\sim10^{15}$. In the RS model this is obtained predominantly with a small value of $f_n^{\prime\prime}(r_{\rm{uv}})$. However an important point is that the required size of $\Omega$ is determined by the relative scaling of the fundamental Planck mass (\ref{PlankMass}). A space of large volume would suppress the fundamental Planck mass and not require as large a warp factor. Assuming for the moment $\Omega$ must be very large. We could, for example, get $F_n\sim 1$ if $f_n$ was close to constant for most of the space but, in order to generate a large warping, the wave function would have to either blow up (or be suppressed) in the UV (IR) while simultaneously the second derivative would have to be small (or large). Alternatively the warping could be achieved through the $b$ term but this would have to be done such that the 4D Planck mass was not enhanced. Of course metrics can always be written down that satisfy these conditions but here we agree with \cite{Delgado:2007ne} in saying that such metrics would be in danger of being contrived.  
 
 \subsection{$F_n$ for a Space with a Varying Radius in the Internal Manifold}
 Returning now to the case of $D$ dimensions it is trivial to see, in (\ref{gaugeEQN}) and (\ref{fnsmall}), that when $c(r)$ is constant there is no deviation in the wavefunctions between the 5D modes and the non harmonic modes. There is also clearly no deviation in the relative gauge-Higgs coupling. Hence the phenomenological implications of considering $D>5$, where the internal manifold has a constant radius, will come predominantly from the degeneracy resulting from the harmonic modes \cite{Davoudiasl:2008qm}.  
 
However here we consider a toy model of a $D$-dimensional space where the radius of the internal space scales as a power law with respect to the warped dimension
\begin{equation}
\label{ }
ds^2=\frac{r^2}{R^2}\eta_{\mu\nu}dx^\mu dx^\nu-\frac{R^2}{r^2}\left (dr^2+\frac{r^{2+a}}{R^a}d\Omega_\delta^2\right ).
\end{equation} 
Clearly if one sets $\delta$ to zero and makes the co-ordinate transformation $r=\exp(-\frac{y}{R})$, then one regains the Randall and Sundrum metric. Further $a=0$ corresponds to the AdS$_5\times\mathcal{M}^\delta$ studied in for example \cite{Davoudiasl:2002wz}. Here we set $r_{\rm{uv}}=R$ and $r_{\rm{ir}}=R^{\prime}$, such that  $R^{\prime}\leqslant r \leqslant R$ and $\Omega=\frac{R}{R^{\prime}}$. However as mentioned before the size of warp factor required to resolve the hierarchy problem is dependent on the scaling of the Planck mass (\ref{PlankMass})
\begin{equation}
\label{ }
M_{\rm{P}}^2\sim\left (\int d^\delta x\sqrt{\gamma}\right )\frac{2}{4+a\delta}R^{\delta+1}\left(1-\left (\frac{1}{\Omega}\right )^{2+\frac{a\delta}{2}}\right )M_{\rm{fund}}^{\delta+3}.
\end{equation}
If we assume that $\left (\int d^\delta x\sqrt{\gamma}\right )$ is of order one and $2+\frac{a\delta}{2}>0$ then when $R\sim M_{\rm{P}}^{-1}$ a warp factor of $\Omega\sim10^{15}$ would resolve the hierarchy problem. The gauge field wavefunctions would then be described by (\ref{gaugeEQN}) as
\begin{equation}
\label{geqnRr}
f_n^{\prime\prime}+\frac{6+a\delta}{2r}f_n^{\prime}-\frac{R^a}{r^{2+a}}\alpha_nf_n+m_n^2\frac{R^4}{r^4}f_n=0.
\end{equation}
If we now consider the non harmonic modes ($\alpha_n=0$ and $\Theta_n=$ constant) then (\ref{geqnRr}) can be solved to give
\begin{equation}
\label{ }
f_n=\frac{N}{r^{1+\frac{a\delta}{4}}}\left (J_{-1-\frac{a\delta}{4}}\left (\frac{R^2m_n}{r}\right )+\beta Y_{-1-\frac{a\delta}{4}}\left (\frac{R^2m_n}{r}\right )\right ).
\end{equation}
If we now define $\hat{m}_n\equiv\frac{R^2m_n}{R^{\prime}}$ then under Neumann BC's 
\begin{displaymath}
\beta=-\frac{J_{-\frac{a\delta}{4}}(\hat{m}_n\Omega^{-1})}{Y_{\frac{-a\delta}{4}}(\hat{m}_n\Omega^{-1})}
\end{displaymath}
and
\begin{displaymath}
J_{-\frac{a\delta}{4}}(\hat{m}_n)+\beta Y_{-\frac{a\delta}{4}}(\hat{m}_n)=0.
\end{displaymath}
The normalisation constant is given by
\begin{displaymath}
N^{-2}=\frac{R^{1+\delta-\frac{a\delta}{2}}}{(R^{\prime}\hat{m}_n)^2}\bigg [\frac{x^2}{2}\bigg (J_v(x)^2-J_{v-1}(x)J_{v+1}(x)+\beta^2\bigg(Y_v(x)^2-Y_{v-1}(x)Y_{v+1}(x)\bigg)
\end{displaymath}
\begin{displaymath}
\hspace{3cm}+\beta\bigg(2Y_v(x)J_v(x)-Y_{v-1}(x)J_{v+1}(x)-J_{v-1}(x)Y_{v+1}(x)\bigg)\bigg )\bigg ]_{\hat{m}_n\Omega^{-1}}^{\hat{m}_n}
\end{displaymath}
Where $v=-1-\frac{a\delta}{4}$. If we assume that $\beta$ is small and that $\Omega$ is large then we can make the approximation
\begin{displaymath}
f_n(R^{\prime})\sim\frac{\sqrt{2}}{\sqrt{R^{1+\delta}}}\Omega^{\frac{a\delta}{4}}.
\end{displaymath}
The zero mode on the other hand is given by
\begin{displaymath}
f_0=\left\{\begin{array}{cc}  \frac{1}{\sqrt{R^{1+\delta}\ln (\Omega)}} & \mbox{for $a=0$ or $\delta=0$} \\
\frac{1}{\sqrt{\frac{2R^{1+\delta}}{a\delta}\left [1-\Omega^{-\frac{a\delta}{2}}\right ]}} & \mbox{otherwise}
\end{array}\right.
\end{displaymath}
Putting this together we find that when $\Omega$ is large and $\beta$ is small then the gauge-Higgs coupling is given by
\begin{equation}
\label{ }
F_n\sim\left\{\begin{array}{cc} \sqrt{2\ln \Omega} & \mbox{for $a=0$ or $\delta=0$} \\
\frac{2}{\sqrt{a\delta}}\Omega^{\frac{a\delta}{4}} & \mbox{otherwise}
\end{array}\right .
\end{equation}

\begin{figure}
\begin{center}
\includegraphics[width=6in,height=3.5in]{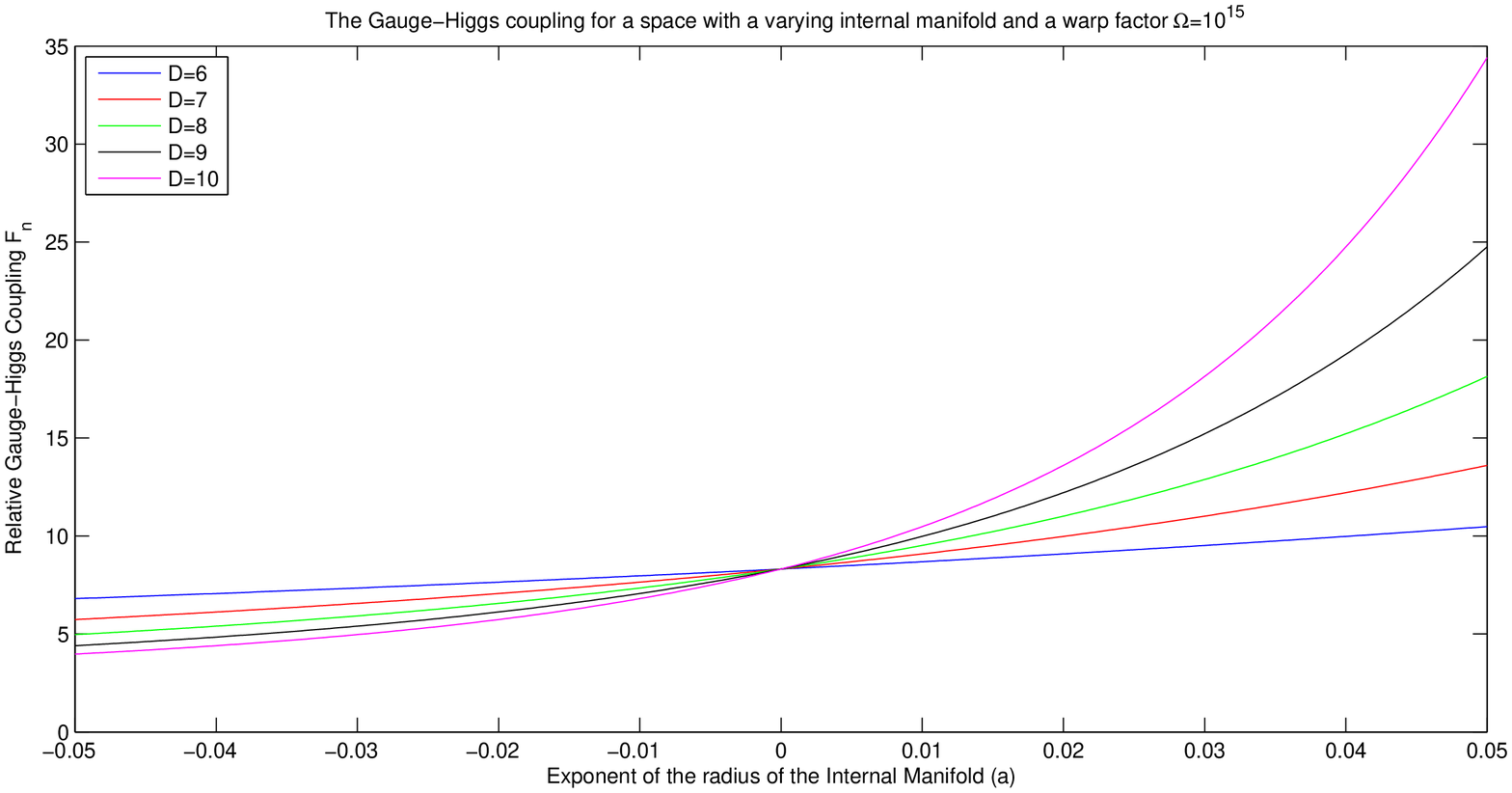}
\caption{The relative gauge Higgs coupling for gauge fields propagating in a $D$-dimensional space with a internal manifold growing as $r^a$. Here $R=10^{-18}$ however in practice the results are not very sensitive to this number.  }
\label{FnRa}
\end{center}
\end{figure}

This is only an approximate result which tends to break down for $\beta\gtrsim-0.2$, but it does emphasize an important result. Notably that in spaces where the size of the internal manifold grows (or shrinks) towards the UV then the gauge-Higgs coupling scales as a function of the warp factor. The exact couplings are shown in figure {\ref{FnRa}}. In the case where the space shrinks towards the UV the coupling asymptotes to one and hence in theory a space could be found that resolves the hierarchy problem and has lower EW constraints than that of UED's. On the other hand where the internal spaces are growing towards the UV the coupling blows up and the KK gauge modes would become strongly coupled. In this case the tree level perturbative EW analysis done here would not be valid but it is reasonable to assume the EW constraints would be large. Although here we have looked at spaces growing as a power law it is straight forward to check that the effect also appears when the space is for example growing exponentially. One should not really be surprised by this result since it is just the volume enhancement / suppression of the coupling as used in the ADD model. Although it is not exactly the same since here it is the relative coupling that is of importance and hence a universal scaling of both the zero mode and KK mode would have no effect. But it is in warped scenarios, where the volume effect in the overlap integral scales the KK modes differently from the zero mode, that we get a phenomenological effect. This effect has also been observed in \cite{McDonald:2009hf}. 

To get an idea of the size of the constraints one can calculate the constraints arising from $s_Z^2$ for a 10D space for three cases corresponding to fermions localised to the IR brane ($F_\psi^{(n)}=F_n$), fermions localised away from the gauge fields ($F_\psi^{(n)}=0.1F_n$) and fermions completely decouple from the gauge fields ($F_\psi^{(n)}=0$). When $a=-0.05$ one obtains the lower bounds on the first gauge mode mass of approximately $14$TeV, $9$TeV and $7$TeV. While when $a=0$ one obtains constraints similar to those of the RS model, $30$TeV, $19$TeV and $15$TeV. However if $a=0.05$ the constraints rise to $127$TeV, $78$TeV and $61$TeV.

Of interest to us now is the question, is it `realistic' to consider spaces with growing internal manifolds? Here we consider the effect of letting gauge fields propagate in a popular solution of type IIb supergravity \cite{Klebanov:2000hb}.

\subsection{The Klebanov-Strassler Solution}
Both AdS$_D$ and AdS$_5\times\mathcal{M}^\delta$ spaces suffer from a conical singularity in the IR and hence in order to consider QFT's propagating in such a background it is necessary to cut the space off with an IR brane. The Klebanov- Strassler solution, corresponding to the dual of $\mathcal{N}=1\quad SU(N+M)\times SU(N)$, is a deformed conifold and hence has no such singularity. Unfortunately it has a quite complex form and hence can only really be investigated numerically. The metric is then  
\begin{equation}
\label{KSmetric }
ds_{10}^2=h^{-\frac{1}{2}}(\tau)\eta_{\mu\nu}dx^\mu dx^\nu - h^{\frac{1}{2}}(\tau)ds_{6}^2
\end{equation} 
where
\begin{eqnarray}
h(\tau)&=&2^{\frac{2}{3}}(g_sM\alpha^{\prime})^2\epsilon^{\frac{-8}{3}}I(\tau) \nonumber\\[.2cm]
I(\tau)&=&\int_{\tau}^{\infty}dx \frac{x\coth x-1}{\sinh^2x}(\sinh(2x)-2x)^{\frac{1}{3}}\nonumber.
\end{eqnarray}
While the internal manifold is described by 
\begin{displaymath}
ds_{6}^2=\frac{1}{2}\epsilon^{\frac{4}{3}}K(\tau)\left [ \frac{1}{3K^3(\tau)}(d\tau^2+(g^5)^2) +\cosh^2(\frac{\tau}{2})[(g^3)^2+(g^4)^2]+\sinh^2(\frac{\tau}{2})[(g^1)^2+(g^2)^2]\right ]
\end{displaymath}
where $g^1$ to  $g^5$ are a diagonal combinations of the angular coordinates, $\epsilon$ is a parameter specifying the conic radius at which deforming begins and
\begin{equation}
\label{Ktau }
K(\tau)=\frac{(\sinh(2\tau)-2\tau)^{\frac{1}{3}}}{2^{\frac{1}{3}}\sinh(\tau)}.\nonumber
\end{equation}
In the IR we assume the Higgs is localised to a point $\tau_{ir}$ close to $\tau=0$ while we still cut the space off at $\tau_{uv}$ such that $h(t_{uv})=1$.
The gauge field propagates in the bulk. For small $\tau$ you can make the approximation $\tau$ $(3K^3(\tau))^{-1}\approx \cosh^2(\tau)$ and write the internal manifold as 
\begin{equation}
\label{ds6approx}
ds_6^2\approx\frac{1}{2}\epsilon^{\frac{4}{3}}K(\tau)\left [\frac{d\tau^2}{3K^3(\tau)} +\cosh^2(\frac{\tau}{2})d\Omega_3^2+\sinh^2(\frac{\tau}{2})d\Omega_2^2\right ], 
\end{equation}
where $d\Omega_3^2$ and $d\Omega_2^2$ are the line elements for $S^3$ and $S^2$. Hence the space is growing towards the UV and so we would anticipate the gauge-Higgs couplings to become large. The non-harmonic KK modes of gauge fields propagating in such a background are then described by (\ref{gaugeEQN})
\begin{equation}
\label{ }
f_n^{\prime\prime}+\left (\frac{h^{\prime}(\tau)}{2h(\tau)}+2\coth(\tau)\right )f_n^{\prime}+\frac{\epsilon h(\tau)}{6K^2(\tau)}m_n^2f_n=0
\end{equation}
Where now the orthogonality relation is given by
\begin{equation}
\label{ }
\int d\tau \frac{h^{\frac{3}{2}}\sinh^2(\tau)\epsilon^4}{24}f_nf_m=\delta_{nm}.
\end{equation}

\begin{figure}
\begin{center}
\includegraphics[width=5in,height=3in]{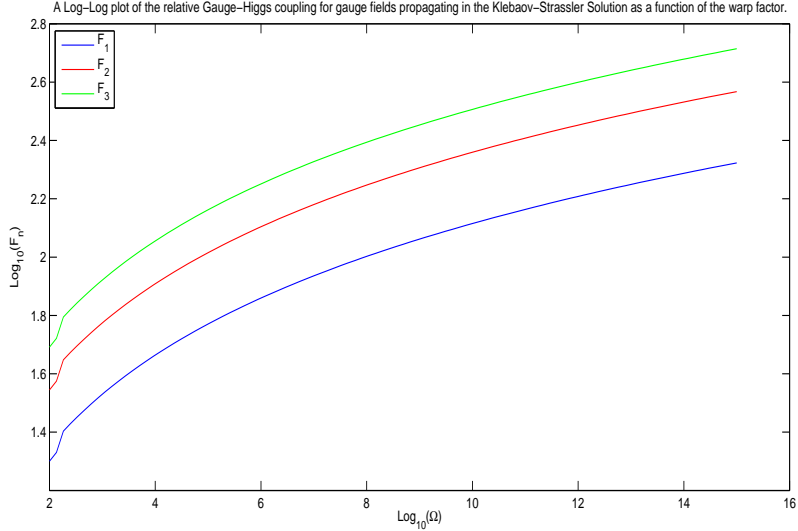}
\caption{The relative gauge Higgs coupling as a function of the warp factor. Here $\epsilon$ is taken to be $10^{-35}$ but the results are not significantly dependent on this value.}
\label{KSlogOm}
\end{center}
\end{figure}

Due to the inaccuracies associated with using splines in differential equations we also approximate 
\begin{displaymath}
I(\tau)\approx\frac{I_0+a\tau^8(\tau-\frac{1}{4})}{1+b\tau^2+c\tau^8e^{\frac{4\tau}{3}}}
\end{displaymath}
where $I_0$, $a$, $b$ and $c$ are choosen so as to fit with the exact function. In practice we take $I_0\approx0.71805$, $a\approx0.03$, $b\approx0.278$ and $c\approx0.0126$. The gauge Higgs couplings are then plotted in figure \ref{KSlogOm}. As expected the relative couplings become large, e.g.~for $\Omega=10^{15}$ we have $F_1\approx210$. Unlike in the RS model the KK modes do not couple with approximately equal strength.

It is also worth noting that for gauge fields propagating in the 5D KS background
\begin{displaymath}
ds^2=ds_{5}^2=h^{-\frac{1}{2}}(\tau)\eta_{\mu\nu}dx^\mu dx^\nu- h^{\frac{1}{2}}(\tau)d\tau^2
\end{displaymath}    
the relative couplings do not shift significantly from those of the RS model. For example for an $\Omega=10^{15}$ gives $F_1\approx7.30$ and $F_2\approx8.05$. Clearly in the Klebanov-Strassler background the main effect on the gauge Higgs coupling is related to the growing size of the internal space, not to the modified warp factor.

 As mentioned before it is probably not meaningful to convert these large couplings into EW constraints using the naive tree level calculation of the previous section. However we can infer that the constraints would be very large and any KK gauge bosons clearly would be outside the reach of LHC. We conclude that bulk gauge fields in a Klebanov-Strassler background do not constitute a viable weak scale extension of the SM.
 
 \section{Discussion and Conclusions}
In this paper we have computed the tree level contribution to EW observables arising from SM particles propagating in a D dimensional spacetime warped with respect to a single extra dimension. It is found that the corrections are largely dependent on three parameters the relative gauge-Higgs coupling, relative gauge-fermion coupling and inversely related to the KK mass scale. In a best case scenario it is assumed that the gauge-fermion coupling could be suppressed by localising the fermion zero mode away from the KK gauge fields. In 5D this is straight forward to do with the inclusion of a bulk Dirac mass. However in $D>5$ it is less clear that the appropriate boundary conditions can be found so as to achieve this and still regain the correct low energy 4D effective chiral theory. An alternative to this scenario would be to localise the fermions on the IR brane in which case the relative gauge fermion coupling would simply be the same as the relative gauge Higgs coupling. A worst case scenario would of course be that the gauge-fermion coupling is larger than the gauge fermion coupling.

In the above three possibilities what is clear is that if a space has a large relative gauge-Higgs coupling then the KK mass scale must be large to fit with existing observations. In 5D we demonstrate that if a space resolves the gauge hierarchy problem using warping, while it is not impossible to have a small gauge Higgs coupling ($F_n\lesssim 1$), it would require a quite contrived space. Hence we agree with \cite{Delgado:2007ne} in saying that large EW constraints are a generic feature of 5D `natural' spaces that resolve the hierarchy problem through warping.

However in D dimensions we find that the relative gauge-Higgs coupling appears to scale as a function of the warp factor. By studying a space in which the radius of the internal space is increasing (decreasing) towards the IR brane, it is found that the gauge-Higgs coupling is significantly suppressed (enhanced). Although this is demonstrated for a space where the internal radius varies as a power law, it is straight forward to demonstrate that it holds for other spaces. This effect is due to the volume enhancement/suppression of the coupling effecting the excited KK modes (in warped spaces) differently than those of the zero mode. Such spaces would of course have significantly raised/lowered EW constraints. This result is tested by computing the gauge-Higgs coupling for fields propagating in the Klebanov Strassler solution \cite{Klebanov:2000hb}. It is found that if the warp factor is taken to be $\Omega=10^{15}$ then the first KK gauge mode couples to the IR brane with the strength of about 210 times that of the zero mode. Hence such a model would not offer a viable resolution to the gauge hierarchy problem.              

Throughout this work the emphasis has been on keeping the model as generic as possible, although it is very difficult to say anything in complete generality, counter examples can nearly always be found. For a start we did not consider the most general metric, it would be interesting to consider the effects of warping in more than one dimension. We also did not complete a thorough study of the fermions, although it is unlikely to change our results if we had. We have also only worked using a naive tree level analysis which is of course not valid when the gauge-Higgs couplings become large, but once again it is hard to see how our results would change had we done a more thorough analysis.

The central point of this paper is that $D>5$ spaces with warped internal manifolds have very different phenomenology to the 5D RS model and clearly more work is needed before we can begin to match experimental data to models.

\section*{Acknowledgements}
P.R.A.~is supported by STFC.

\bibliographystyle{JHEP}

\bibliography{generalwarp}

\providecommand{\href}[2]{#2}\begingroup\raggedright\begin{thebibliography}{10}

\bibitem{Randall:1999ee}
L.~Randall and R.~Sundrum, {\it {A large mass hierarchy from a small extra
  dimension}},  {\em Phys. Rev. Lett.} {\bf 83} (1999) 3370--3373,
  [\href{http://arxiv.org/abs/hep-ph/9905221}{{\tt hep-ph/9905221}}].

\bibitem{Davoudiasl:1999tf}
H.~Davoudiasl, J.~L. Hewett, and T.~G. Rizzo, {\it {Bulk gauge fields in the
  Randall-Sundrum model}},  {\em Phys. Lett.} {\bf B473} (2000) 43--49,
  [\href{http://arxiv.org/abs/hep-ph/9911262}{{\tt hep-ph/9911262}}].

\bibitem{Pomarol:1999ad}
A.~Pomarol, {\it {Gauge bosons in a five-dimensional theory with localized
  gravity}},  {\em Phys. Lett.} {\bf B486} (2000) 153--157,
  [\href{http://arxiv.org/abs/hep-ph/9911294}{{\tt hep-ph/9911294}}].

\bibitem{Grossman:1999ra}
Y.~Grossman and M.~Neubert, {\it {Neutrino masses and mixings in
  non-factorizable geometry}},  {\em Phys. Lett.} {\bf B474} (2000) 361--371,
  [\href{http://arxiv.org/abs/hep-ph/9912408}{{\tt hep-ph/9912408}}].

\bibitem{Chang:1999nh}
S.~Chang, J.~Hisano, H.~Nakano, N.~Okada, and M.~Yamaguchi, {\it {Bulk standard
  model in the Randall-Sundrum background}},  {\em Phys. Rev.} {\bf D62} (2000)
  084025, [\href{http://arxiv.org/abs/hep-ph/9912498}{{\tt hep-ph/9912498}}].

\bibitem{Huber:2000ie}
S.~J. Huber and Q.~Shafi, {\it {Fermion Masses, Mixings and Proton Decay in a
  Randall- Sundrum Model}},  {\em Phys. Lett.} {\bf B498} (2001) 256--262,
  [\href{http://arxiv.org/abs/hep-ph/0010195}{{\tt hep-ph/0010195}}].

\bibitem{Gherghetta:2000qt}
T.~Gherghetta and A.~Pomarol, {\it {Bulk fields and supersymmetry in a slice of
  AdS}},  {\em Nucl. Phys.} {\bf B586} (2000) 141--162,
  [\href{http://arxiv.org/abs/hep-ph/0003129}{{\tt hep-ph/0003129}}].

\bibitem{Agashe:2004cp}
K.~Agashe, G.~Perez, and A.~Soni, {\it {Flavor structure of warped extra
  dimension models}},  {\em Phys. Rev.} {\bf D71} (2005) 016002,
  [\href{http://arxiv.org/abs/hep-ph/0408134}{{\tt hep-ph/0408134}}].

\bibitem{ArkaniHamed:2000ds}
N.~Arkani-Hamed, M.~Porrati, and L.~Randall, {\it {Holography and
  phenomenology}},  {\em JHEP} {\bf 08} (2001) 017,
  [\href{http://arxiv.org/abs/hep-th/0012148}{{\tt hep-th/0012148}}].

\bibitem{Maldacena:1997re}
J.~M. Maldacena, {\it {The large N limit of superconformal field theories and
  supergravity}},  {\em Adv. Theor. Math. Phys.} {\bf 2} (1998) 231--252,
  [\href{http://arxiv.org/abs/hep-th/9711200}{{\tt hep-th/9711200}}].

\bibitem{Klebanov:2000nc}
I.~R. Klebanov and A.~A. Tseytlin, {\it {Gravity duals of supersymmetric SU(N)
  x SU(N+M) gauge theories}},  {\em Nucl. Phys.} {\bf B578} (2000) 123--138,
  [\href{http://arxiv.org/abs/hep-th/0002159}{{\tt hep-th/0002159}}].

\bibitem{Klebanov:2000hb}
I.~R. Klebanov and M.~J. Strassler, {\it {Supergravity and a confining gauge
  theory: Duality cascades and chiSB-resolution of naked singularities}},  {\em
  JHEP} {\bf 08} (2000) 052, [\href{http://arxiv.org/abs/hep-th/0007191}{{\tt
  hep-th/0007191}}].

\bibitem{Csaki:2002gy}
C.~Csaki, J.~Erlich, and J.~Terning, {\it {The effective Lagrangian in the
  Randall-Sundrum model and electroweak physics}},  {\em Phys. Rev.} {\bf D66}
  (2002) 064021, [\href{http://arxiv.org/abs/hep-ph/0203034}{{\tt
  hep-ph/0203034}}].

\bibitem{carena-2003-68}
M.~Carena, A.~Delgado, E.~Ponton, T.~M.~P. Tait, and C.~E.~M. Wagner, {\it
  Precision electroweak data and unification of couplings in warped extra
  dimensions},  {\em Physical Review D} {\bf 68} (2003) 035010.

\bibitem{Agashe:2003zs}
K.~Agashe, A.~Delgado, M.~J. May, and R.~Sundrum, {\it {RS1, custodial isospin
  and precision tests}},  {\em JHEP} {\bf 08} (2003) 050,
  [\href{http://arxiv.org/abs/hep-ph/0308036}{{\tt hep-ph/0308036}}].

\bibitem{Davoudiasl:2002wz}
H.~Davoudiasl, J.~L. Hewett, and T.~G. Rizzo, {\it {Phenomenology on a slice of
  AdS(5) $\times$ $M^{\delta}$ spacetime}},  {\em JHEP} {\bf 04} (2003) 001,
  [\href{http://arxiv.org/abs/hep-ph/0211377}{{\tt hep-ph/0211377}}].

\bibitem{Davoudiasl:2008qm}
H.~Davoudiasl and T.~G. Rizzo, {\it {New Dimensions for Randall-Sundrum
  Phenomenology}},  {\em JHEP} {\bf 11} (2008) 013,
  [\href{http://arxiv.org/abs/0809.4440}{{\tt arXiv:0809.4440}}].

\bibitem{Delgado:2007ne}
A.~Delgado and A.~Falkowski, {\it {Electroweak observables in a general 5D
  background}},  {\em JHEP} {\bf 05} (2007) 097,
  [\href{http://arxiv.org/abs/hep-ph/0702234}{{\tt hep-ph/0702234}}].

\bibitem{McGuirk:2007er}
P.~McGuirk, G.~Shiu, and K.~M. Zurek, {\it {Phenomenology of Infrared Smooth
  Warped Extra Dimensions}},  {\em JHEP} {\bf 03} (2008) 012,
  [\href{http://arxiv.org/abs/0712.2264}{{\tt arXiv:0712.2264}}].

\bibitem{McDonald:2009hf}
K.~L. McDonald, {\it {Warping, Extra Dimensions and a Slice of $AdS_d$}},  {\em
  Phys. Rev.} {\bf D81} (2010) 024006,
  [\href{http://arxiv.org/abs/0909.5454}{{\tt arXiv:0909.5454}}].

\bibitem{McDonald:2009md}
K.~L. McDonald, {\it {Warping the Universal Extra Dimensions}},  {\em Phys.
  Rev.} {\bf D80} (2009) 024038, [\href{http://arxiv.org/abs/0905.3006}{{\tt
  arXiv:0905.3006}}].

\bibitem{ArkaniHamed:1998rs}
N.~Arkani-Hamed, S.~Dimopoulos, and G.~R. Dvali, {\it {The hierarchy problem
  and new dimensions at a millimeter}},  {\em Phys. Lett.} {\bf B429} (1998)
  263--272, [\href{http://arxiv.org/abs/hep-ph/9803315}{{\tt hep-ph/9803315}}].

\bibitem{Antoniadis:1998ig}
I.~Antoniadis, N.~Arkani-Hamed, S.~Dimopoulos, and G.~R. Dvali, {\it {New
  dimensions at a millimeter to a Fermi and superstrings at a TeV}},  {\em
  Phys. Lett.} {\bf B436} (1998) 257--263,
  [\href{http://arxiv.org/abs/hep-ph/9804398}{{\tt hep-ph/9804398}}].

\bibitem{Davoudiasl:2008hx}
H.~Davoudiasl, G.~Perez, and A.~Soni, {\it {The Little Randall-Sundrum Model at
  the Large Hadron Collider}},  {\em Phys. Lett.} {\bf B665} (2008) 67--71,
  [\href{http://arxiv.org/abs/0802.0203}{{\tt arXiv:0802.0203}}].

\bibitem{Burdman:2005sr}
G.~Burdman, B.~A. Dobrescu, and E.~Ponton, {\it {Six-dimensional gauge theory
  on the chiral square}},  {\em JHEP} {\bf 02} (2006) 033,
  [\href{http://arxiv.org/abs/hep-ph/0506334}{{\tt hep-ph/0506334}}].

\bibitem{Peskin:1991sw}
M.~E. Peskin and T.~Takeuchi, {\it {Estimation of oblique electroweak
  corrections}},  {\em Phys. Rev.} {\bf D46} (1992) 381--409.

\bibitem{Amsler:2008zzb}
{\bf Particle Data Group} Collaboration, C.~Amsler {\em et~al.}, {\it {Review
  of particle physics}},  {\em Phys. Lett.} {\bf B667} (2008) 1.

\bibitem{Huber:2001gw}
S.~J. Huber, C.-A. Lee, and Q.~Shafi, {\it {Kaluza-Klein excitations of W and Z
  at the LHC?}},  {\em Phys. Lett.} {\bf B531} (2002) 112--118,
  [\href{http://arxiv.org/abs/hep-ph/0111465}{{\tt hep-ph/0111465}}].

\bibitem{Goertz:2008vr}
F.~Goertz and T.~Pfoh, {\it {On the Perturbative Approach in the
  Randall-Sundrum Model}},  {\em JHEP} {\bf 10} (2008) 035,
  [\href{http://arxiv.org/abs/0809.1378}{{\tt arXiv:0809.1378}}].

\bibitem{Aaltonen:2007ps}
{\bf CDF} Collaboration, T.~Aaltonen {\em et~al.}, {\it {First Run II
  Measurement of the $W$ Boson Mass}},  {\em Phys. Rev.} {\bf D77} (2008)
  112001, [\href{http://arxiv.org/abs/0708.3642}{{\tt arXiv:0708.3642}}].

\end{thebibliography}\endgroup

\end{document}